\documentclass[twocolumn,3p]{elsarticle}
\usepackage[latin9]{inputenc}
\usepackage[english]{babel}
\usepackage{verbatim}
\usepackage{amsthm}
\usepackage{hyperref}
\usepackage{algorithm}%
\usepackage{algpseudocode}%
\usepackage{algorithmicx}%
\usepackage{listings}%
\usepackage{paralist}%
\usepackage{xspace}
\usepackage{xcolor}%
\usepackage{amsmath,amsfonts}%
\usepackage{booktabs}%
\usepackage{tikz}
\usepackage{enumitem}
\usepackage{eqparbox}
\usepackage{pgf}
\usepackage{tikz}
\usetikzlibrary{arrows,petri}
\usepackage[normalem]{ulem}

\newtheorem{example}{Example}%
\newtheorem{definition}{Definition}%

\newcommand\StateX{\Statex\hspace{\algorithmicindent}}

\newcommand{\tool}{\textsc{BenCIGen}\xspace}
\def\checkmark{\tikz\fill[scale=0.4](0,.35) -- (.25,0) -- (1,.7) -- (.25,.15) -- cycle;}

\newcommand{\ins}[1]{#1} 
\newcommand{\del}[1]{} 
\newcommand{\chg}[2]{#2} 

\newdimen{\algindent}
\algnewcommand\LeftComment[2]{%
\hspace{#1\algindent}$\triangleright$ \eqparbox{COMMENT}{#2} \hfill %
}
\algnewcommand\And{\textbf{and} }
\algnewcommand\Not{\textbf{not} }
\algnewcommand\Or{\textbf{or} }
\algnewcommand{\IIf}[1]{\State\algorithmicif\ #1\ \algorithmicthen}
\algnewcommand{\EndIIf}{\unskip\ \algorithmicend\ \algorithmicif}

\lstdefinelanguage{CTWedge} 
{morekeywords={Model, Parameters, Constraints},
	sensitive=true, morecomment=[l]{//}, morecomment=[s]{/*}{*/},
	morecomment=[l][\color{white}\tiny]{'},
	morestring=[b]",tabsize=2, columns=fullflexible, basicstyle=\sffamily\small, 
	captionpos=b, breaklines=true, frame=single}

\colorlet{punct}{red!60!black}
\definecolor{background}{HTML}{EEEEEE}
\definecolor{delim}{RGB}{20,105,176}
\colorlet{numb}{magenta!60!black}

\lstdefinelanguage{json}{
	basicstyle=\normalfont\ttfamily,
	showstringspaces=false,
	breaklines=true,
	frame=single,
	tabsize=2,
	literate=
	*{0}{{{\color{numb}0}}}{1}
	{1}{{{\color{numb}1}}}{1}
	{2}{{{\color{numb}2}}}{1}
	{3}{{{\color{numb}3}}}{1}
	{4}{{{\color{numb}4}}}{1}
	{5}{{{\color{numb}5}}}{1}
	{6}{{{\color{numb}6}}}{1}
	{7}{{{\color{numb}7}}}{1}
	{8}{{{\color{numb}8}}}{1}
	{9}{{{\color{numb}9}}}{1}
	{:}{{{\color{punct}{:}}}}{1}
	{,}{{{\color{punct}{,}}}}{1}
	{\{}{{{\color{delim}{\{}}}}{1}
	{\}}{{{\color{delim}{\}}}}}{1}
	{[}{{{\color{delim}{[}}}}{1}
	{]}{{{\color{delim}{]}}}}{1},
}

\journal{The Journal of Systems \& Software}

\begin{document}
	
\begin{frontmatter}{}

\title{Design, implementation, and validation of a benchmark generator for combinatorial interaction testing tools}

\author[rvt]{Andrea Bombarda\corref{cor1}}

\ead{andrea.bombarda@unibg.it}

\author[rvt]{Angelo Gargantini\corref{cor1}}

\ead{angelo.gargantini@unibg.it}

\cortext[cor1]{Corresponding author}

\address[rvt]{Department of Engineering, University of Bergamo, Bergamo, Italy}

\begin{abstract}
	
Combinatorial testing is a widely adopted technique for efficiently detecting faults in software. 
The quality of combinatorial test generators plays a crucial role in achieving effective test coverage.
Evaluating combinatorial test generators remains a challenging task that requires 
diverse and representative benchmarks.
Having such benchmarks might help developers to test their tools, and improve their performance.
	
For this reason, in this paper, we present \tool, a highly configurable generator of benchmarks to be used by combinatorial test generators, empowering users to customize the type of benchmarks generated, including constraints and parameters, as well as their complexity.
An initial version of such a tool has been used during 
the CT-Competition, held yearly during the International Workshop on Combinatorial Testing. 
This paper describes the requirements, the design, the implementation, and the validation of \tool.  
Tests for the validation of \tool are derived from its requirements by using a combinatorial interaction approach.
Moreover, we demonstrate the tool's ability to generate benchmarks that reflect the characteristics of real software systems. 
	
\tool not only facilitates the evaluation of existing generators but also serves as a valuable resource for researchers and practitioners seeking to enhance the quality and effectiveness of combinatorial testing methodologies.

\end{abstract}




\begin{keyword}
combinatorial testing \sep benchmarks \sep test generators \sep validation
\end{keyword}

\end{frontmatter}{}

\section{Introduction}\label{sec:introduction}

Combinatorial Interaction Testing (CIT)~\cite{petke2015practical} has been an active area of research in the latest years and has proven to be very effective to test complex systems, having multiple inputs or configuration parameters.
The main purpose of CIT is to help testers in finding defects due to the interaction of different inputs or parameters, by testing this interaction systematically and by assuring that every $t$-uple of parameter values \ins{(i.e., an array of $t$ elements, where each element is one of the parameters of the system under test with one of its possible values~\cite{Niu2013})} is tested at least once~\cite{Kuhn2004}.
In practice, testers provide an input parameter model (IPM) of a system under test (SUT), containing the possible values for each parameter, as well as any additional constraints between values of distinct parameters, and ask a test generator to produce a test suite.

During the years, several test generators have been proposed\footnote{For a non-exhaustive list of tools see, for instance, \url{https://www.pairwise.org/tools.html}.} by the community: research groups that actively work on the CIT area have been listed in~\cite{Nie2011}, but many other recent groups and tools are not considered in that paper, while in~\cite{Khalsa2014} a lot of algorithms and tools available for CIT are analyzed. 
\chg{However, paradoxically, the main drawback when developing CIT test generators, which are used to generate tests for other systems, is the absence of a collection of benchmarks to be used for testing the correctness and evaluating the performance of the generators themselves.}{Despite so many algorithms and tools for CIT have been developed with the intent of improving testing of software systems, paradoxically little attention has been given to testing and systematically and fairly evaluating those tools and algorithms. One major issue is the absence of a collection of benchmarks to be used for testing the correctness and evaluating the performance of the generators themselves. Many tools have only been evaluated on ad-hoc or unrealistic models, or small examples, missing some important and common problem characteristics.}
This becomes especially evident when dealing with problems that involve constraints, as they pose a greater challenge for test generators, and obtaining representative test IPMs from real scenarios can be difficult.

\chg{For this reason, w}{W}hile evaluating CIT test generators, every research group has established its own procedure and benchmarks, and this can be limiting for many reasons:
\begin{inparaenum}[a)]
	\item some specific features, which may be common in practice, are not considered while testing the test generator;
	\item on the contrary, uncommon features may be considered and, thus, bias the test outcome;
	\item a limited amount of test IPMs may be available.
\end{inparaenum}

Moreover, having a high number of benchmarks may foster the improvement of the performance of test generators, since they can be tested (and, thus, adapted) against different IPMs.
This is the rationale behind the CT-Competition which is organized every year during the International Workshop of Combinatorial Testing\footnote{\url{https://fmselab.github.io/ct-competition/}}.

To address all these issues, in this paper, we present \tool, a benchmark generator of IPMs that can be used by practitioners to generate synthetic IPMs for testing CIT generators.
First, we design \tool by building a feature model describing its configuration parameters  and possible constraints among them.
\tool is built on the top of the CTWedge environment~\cite{Bombarda2021}, and allows practitioners to generate a set of different benchmarks, with a configurable type, amount, and cardinality of parameters and constraints. 
In order to make the benchmarks as challenging as desired, \tool allows users to configure the ratio of the generated IPMs, i.e., the fraction of the number of valid tests (or $t$-uples) over the total number of tests (or $t$-uples).
We believe that this aspect is crucial for assessing the performance of a test generator under different (also in terms of complexity) use case scenarios.
Lastly, \tool only produces \chg{valid}{solvable} IPMs, i.e., IPMs from which at least a test case can be generated.
This is of paramount importance for making the use of generated benchmarks valuable for evaluating test generators: assessing the performance (time and test suite size~\cite{Bombarda2021}) of test generators requires models that allow at least a test case.  
\chg{Invalid}{Non solvable} models could be useful as well in order to test the correctness of test generators \ins{but not in evaluating their performance}, and we may add the feature of generating also \chg{invalid}{non solvable} IPMs in future releases of \tool.


We investigate the correctness of \tool by using combinatorial test cases derived from its model, and we show how models available in the literature can fit inside those that can be generated from our tool.
By demonstrating this aspect, we can state that \tool can generate realistic IPMs, as challenging and complex as those used in practice for real systems, and, thus, that the benchmarks we generate are valuable for effectively testing CIT test generators.

The remainder of the paper is structured as follows.
Sect.~\ref{sec:background} describes the background on combinatorial testing and the measures we perform on each generated IPM.
In Sect.~\ref{sec:requirements} we present the requirements we set for the development of \tool, while Sect.~\ref{sec:deisign} introduces the design of our tool and the possible approaches for computing the two types of ratio and the \chg{validity}{solvability} of an IPM.
Sect.~\ref{sec:implementation} shows \tool and how we have implemented it, while in Sect.~\ref{sec:validation} we validate our tool by generating combinatorial tests from its requirements, and by showing how the majority of CIT models available in the literature can fit in those that our tool can generate.
Finally, Sect.~\ref{sec:related} presents related works on benchmarking combinatorial test generators, and Sect.~\ref{sec:conclusions} concludes the paper.

\section{Background}\label{sec:background}

Combinatorial test generators are tools used to generate test suites suitable for testing a system that has been modeled using an Input Parameter Model (IPM). 
It specifies parameters of a system under test (SUT), their possible values, as well as any additional constraints between values of distinct parameters.
Formally, it can be defined as follows.

\smallskip
\begin{definition}[Input Parameter Model]\label{def:IPM}
	Let $S$ be the system under test, $P=\{p_1, ..., p_n\}$ be a set of $n$ parameters, where every parameter $p_i$ assumes values in the domain $D_i = \{v_1^i, \ldots, v_{j}^i\}$, let $D$ be the set of all the $D_i$, i.e., $D=\{D_1,\ldots, D_k\}$ and $C=\{c_1,...,c_m\}$ be the set of constraints over the parameters $p_i$ and their values $v_j^i$. 
	We say that $M=(P,D,C)$ is an \emph{Input Parameter Model} for the system $S$.
\end{definition}

\smallskip
Given an IPM, test generators build a \emph{test suite} $TS$, composed of several test cases $tc_i$, in which every parameter $p \in P$ has its own value.
The main objective of a $TS$ is to cover all the feasible interactions between $t$ parameters, where $t$ is the strength of the test suite.

\smallskip
\begin{definition}[T-wise coverage]\label{def:twise}
Let $TS$ be the test suite for the IPM $M=(P,D,C)$, as defined in Def.~\ref{def:IPM}, and be $t$ its strength.
We say that $TS$ achieves the \emph{t-wise coverage} if all the feasible $t$-uples among the parameters $p_i \in P$ and their values are covered by at least a test case in $TS$.
\end{definition}

\smallskip
Based on the system to be modeled, the parameters may be of different types. In the work presented in this paper, we consider \emph{Boolean} parameters, that can assume only the true and false values, \emph{Enumerative} parameters, assuming values in a finite set, and \emph{Integer ranges} parameters, assuming values between a lower and an upper bound (both Integers).

An example of IPM, in the CTWedge format~\cite{Gargantini2018}, is given in Listing~\ref{lst:exModel}.
It contains two enumerative parameters (P1 and P3), a single Boolean parameter (P2), and an integer range parameter (P4).
Furthermore, it contains a set of three constraints, defined over the set of parameters.
\ins{Tab.~\ref{tab:testSuiteExample} shows the test suite achieving the pairwise (i.e., $t=2$) coverage for the IPM in Listing~\ref{lst:exModel}.}

\begin{lstlisting}[language=CTWedge, basicstyle=\sffamily\scriptsize, caption=Example of a constrained combinatorial model,float, label=lst:exModel]
	Model example1
	
	Parameters:
	P1 : {V1, V2}
	P2 : Boolean
	P3 : {V1, V2, V3}
	P4 : [2 .. 5]
	
	Constraints:
	# P1 != P3 #
	# (P3=V1 => P2=false) AND P1=V2 #
	# (P4=3 <=> P2=true) OR P3=V3 #
\end{lstlisting}

\begin{table}[]
	\centering
	\caption{\ins{An example of a pairwise ($t=2$) test suite}}
	\label{tab:testSuiteExample}
	\begin{tabular}{@{}cccc@{}}
		\toprule
		\textbf{P1} & \textbf{P2} & \textbf{P3} & \textbf{P4} \\ \midrule
		V2 & false & V1 & 2 \\
		V2 & true  & V3 & 2 \\
		V2 & false & V3 & 3 \\
		V2 & false & V1 & 4 \\
		V2 & true  & V3 & 4 \\
		V2 & false & V1 & 5 \\
		V2 & true  & V3 & 5 \\
		V2 & true  & V3 & 3 \\ \bottomrule
	\end{tabular}
\end{table}

In every IPM, for each constraint, it is possible to compute a complexity, which roughly measures the effort required by the combinatorial test generator when checking the satisfiability of the constraint.
Formally, it can be defined as follows.

\smallskip
\begin{definition}[Complexity]\label{def:complexity}
	Let $M=(P,D,C)$ be an IPM as defined in Def.~\ref{def:IPM}. 
	\ins{The \emph{Complexity} of a constraint $c \in C$ is the number of binary logical operators and connectors in $c$, i.e., the number of AND, OR, implies (\lstinline[language=CTWedge]|=>|), and double implies  (\lstinline[language=CTWedge]|<=>|).} 
	\chg{The \emph{Complexity} of a constraint $c \in C$ can be extracted through the function $Compl : C \rightarrow \mathbb{N}$ that computes the number of logical connectors (AND, OR, implies, and double implies) in $c$.}{More formally, the complexity is represented by a function $Comp : C \rightarrow \mathbb{N}$.}
\end{definition}

\ins{For example, the complexity of the constraint 
\begin{center}
	\lstinline[language=CTWedge]|# P1 = true AND P2 = false #|
\end{center}
is equals to 1, as only a binary logical operator or connector (i.e., the AND) is available. Instead, if we consider the constraint
\begin{center}
	\lstinline[language=CTWedge]|# P1 => (P2 AND P3) #|
\end{center}
the complexity is 2, as we have an AND connector and an implication.
}

\medskip
Given a strength $t$, some of the $t$-uples may clash with one or a conjunction of constraints (i.e., the assignments contained in the $t$-uple violate at least a constraint or a combination of them). 
%
%
%
In that case, none of the tests generated from an IPM will cover those $t$-uples and we say that they are \emph{not feasible} or \emph{invalid}.
In order to measure the effort required to a test generator to filter the not feasible $t$-uples out, we introduce the concept of \emph{Tuple validity ratio} ($r_{tp}$), defined as follows.

\smallskip
\begin{definition}[Tuple validity ratio]\label{def:TupleValidity}
Let $M=(P,D,C)$ be the IPM for a system $S$ and $t$ be the required strength for test generation.
We say that the \emph{tuple validity ratio} $r_{tp}$ is the fraction of the number of valid $t$-uples over the total number of $t$-uples.
\end{definition}

\smallskip
Similarly, due to the constraints, some of the tests that can be generated from an IPM by a combinatorial test generator may be not valid, i.e., they may violate one or more constraints\ins{; instead, tests complying with the constraints of the IPM are considered as \emph{valid}}.
For this reason, to estimate how difficult may be for a generator to generate valid test cases, we exploit the concept of \emph{Test validity ratio} ($r_{ts}$).%

\smallskip
\begin{definition}[Test validity ratio]\label{def:Testvalidity}
Let $M=(P,D,C)$ be the IPM for a system $S$, $TS$ be the set of all possible test cases that can be generated when the constraints $C$ of $M$ are ignored.
Let $TS_v \subseteq TS$ be the set of valid test cases, i.e., the set of those that do not violate any of the constraints in $C$.
We say that the \emph{test validity ratio} $r_{ts}$ is the fraction of the number of valid tests (i.e., the cardinality of $TS_v$) over the total number of possible tests $N$ (i.e., the cardinality of $TS$).
\end{definition}


\section{Requirements}\label{sec:requirements}

During the development of \tool, we aimed at creating a tool allowing users to generate a wide spectrum of IPMs, by specifying all the features and characteristics we have found in the other models available in the literature (see Sect.~\ref{sec:external}).

The possible configurations we wanted to include in \tool generator are reported in the feature model in Fig.~\ref{fig:fm}.
In the following, we better describe the features and their meaning in detail:
\begin{figure*}
	\centering
	\includegraphics[width=\linewidth]{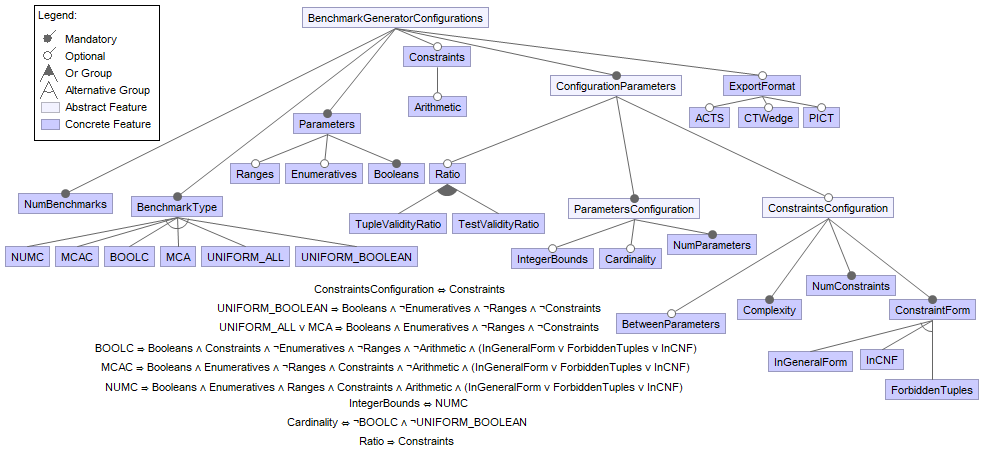}
	\caption{The Feature Model representing the possible configurations and features of the benchmarks generator}
	\label{fig:fm}
\end{figure*}
\begin{itemize}[leftmargin=*]
\item Each generation run may generate multiple benchmarks with the same characteristics. The number of benchmarks is configurable through the feature \textsf{NumBenchmarks};

\item Several different categories (\textsf{BenchmarkType}) of models may be generated, depending on the type of parameters and constraints, as reported in Tab.~\ref{tab:tracks}. In case one of the categories containing constraints is chosen, the \textsf{Constraints} may be selected and, possibly, contain \textsf{Arithmetic} operators;

\item Depending on the benchmark type, different types of \textsf{Parameters} (\textsf{Booleans}, \textsf{Enumeratives}, or integer \textsf{Ranges}) can be present in the generated IPM;

\item Depending on the selected benchmark type, the user may specify the following \textsf{ConfigurationParameters}:
\begin{itemize}[leftmargin=*]
	\item the maximum accepted \textsf{Ratio}, as described in Sect.~\ref{sec:background}, which can be set as the \textsf{TupleValidityRatio} and/or \textsf{TestValidityRatio};
	
	\item regarding the parameters (\textsf{Parameters} \textsf{Configuration}), the user can select:
	\begin{itemize}[leftmargin=*]
		\item the \textsf{Cardinality}, limited between a lower and upper bound, for the parameters in the generated IPMs, only if not BOOLC \chg{nor}{neither} UNIFORM\_BOOLEAN are selected;
		\item the integer ranges (\textsf{IntegerBounds}) in the models of the NUMC category;
		\item the number of parameters \textsf{NumParams} to be present in the generated IPMs, included between a lower and upper bound;
	\end{itemize}
	
	\item regarding the constraints (\textsf{Constraints} \textsf{Configuration}), the user can select:
	\begin{itemize}	[leftmargin=*]
		\item the number of constraints \textsf{NumConstraints} (whether applicable), included between a lower and upper bound;
		\item the \textsf{Complexity}, included between a lower and upper bound, for the constraints in the generated IPMs, as described in Sect.~\ref{sec:background};
		\item whether to have constraints comparison \textsf{BetweenParameters} (e.g., \lstinline[language=CTWedge]|PAR1 = PAR2|) and not only comparisons between parameters and values (e.g., \lstinline[language=CTWedge]|PAR1 = true|);
		\item whether the constraints (if they are applicable - see Tab.~\ref{tab:tracks}) need to be \textsf{InCNF}\footnote{\ins{We support constraints in CNF as some generator, such as in the case of CASA~\cite{Garvin2009}, may require constraints to be defined in that form.}}, expressed as \textsf{ForbiddenTuples} or \textsf{InGeneralForm}. In the first case, each constraint is a conjunction (an AND) of one or more clauses, where a clause is a disjunction (an OR) of \chg{literals}{atomic predicates}. In the second case, each constraint must express a forbidden tuple, i.e., in the form of \lstinline[language=CTWedge]|NOT (P1=v1 AND P2=v2 AND ...)|, or \lstinline[language=CTWedge]|(P1!=v1 OR P2!=v2 OR ...)|. 
		Finally, in the third, \chg{a general}{an arbitrary} composition of each constraint is allowed\ins{, i.e., a mix between conjunctions, disjunctions, implications, equivalences and negations can be used in any arbitrary order and combination};
	\end{itemize}
\end{itemize}
	
\item The generated benchmarks may be exported in different formats, such as \textsf{ACTS}~\cite{Yu2013}, \textsf{PICT}~\cite{PICTRepo} and \textsf{CTWedge}~\cite{Gargantini2018}. \ins{We decided to support these three different formats because they are the most used ones and, moreover, they allow for representing the same type of constraints and operators. Other formats, such as the CASA one, would require the transformation of the constraints and this would make the benchmarks not comparable.}		
\end{itemize}
\begin{table*}
	\centering
	\caption{Types of benchmarks supported by the \tool benchmark generator}
	\label{tab:tracks}
	\begin{tabular}{p{0.24\linewidth}p{0.22\linewidth}p{0.39\linewidth}}
		\toprule
		\textbf{Benchmark Type} & \textbf{Parameters} & \textbf{Constraints}\\
		\midrule	
		\texttt{UNIFORM\_BOOLEAN (UB)} & Only Booleans & NO \\ 	
		\texttt{UNIFORM\_ALL (UA)} & Uniform & NO \\ 		
		\texttt{MCA (M)} & MCA (Booleans and Enumeratives) & NO \\ 
		\texttt{BOOLC (BC)} & Only Booleans & Randomly chosen between \texttt{AND}, \texttt{OR}, $\Leftrightarrow$, \texttt{NOT}, $\Rightarrow$ \\
		\texttt{MCAC (MC)} & MCA (Booleans and Enumeratives) & Randomly chosen between \texttt{AND}, \texttt{OR}, $\Leftrightarrow$, \texttt{NOT}, $\Rightarrow$, $=$ (both $x=C$ and $x=y$, where $x$ and $y$ are parameters and $C$ a constant of $x$), $\neq$ \\
		\texttt{NUMC (NC)} & Booleans, Enums and Integer ranges & Randomly chosen between \texttt{AND}, \texttt{OR}, $\Leftrightarrow$, \texttt{NOT}, $\Rightarrow$, $=$ (both $x=C$ and $x=y$, where $x$ and $y$ are parameters and $C$ a constant of $x$), $\neq$, mathematical and relational operations \\
		\bottomrule
	\end{tabular}
\end{table*}

All these configuration parameters may be set by the user prior to the benchmark generation.
Moreover, considering that in real scenarios one may want to test its combinatorial test generator with models similar to those he/she already has, \tool must provide an interface for extracting the configuration from a former IPM and generating models having similar characteristics.
\ins{Finally, \tool shall allow users to load a JSON file, such as the one in Listing~\ref{lst:dictionary}, representing the dictionary of parameter name, type, and values to be used in the randomly created IPMs when tests of a specific domain are required.}
\begin{figure}[tb]
\begin{lstlisting}[language=json, basicstyle=\sffamily\scriptsize, caption=\ins{Example of a dictionary JSON file for the Smartphone domain for \tool}, label=lst:dictionary]
[
	{
		"name": "ScreenSizeInch",
		"type": "Integer",
		"lowerBound": 4,
		"upperBound": 7
	},
	{
		"name": "OS",
		"type": "Enum",
		"values" : [
			"android",
			"ios"
		]
	},
	{
		"name": "WirelessCharge",
		"type": "Boolean"
	}
]
\end{lstlisting}
\end{figure}

\section{Design}\label{sec:deisign}

\begin{figure}
	\centering
	\includegraphics[width=1.0\linewidth]{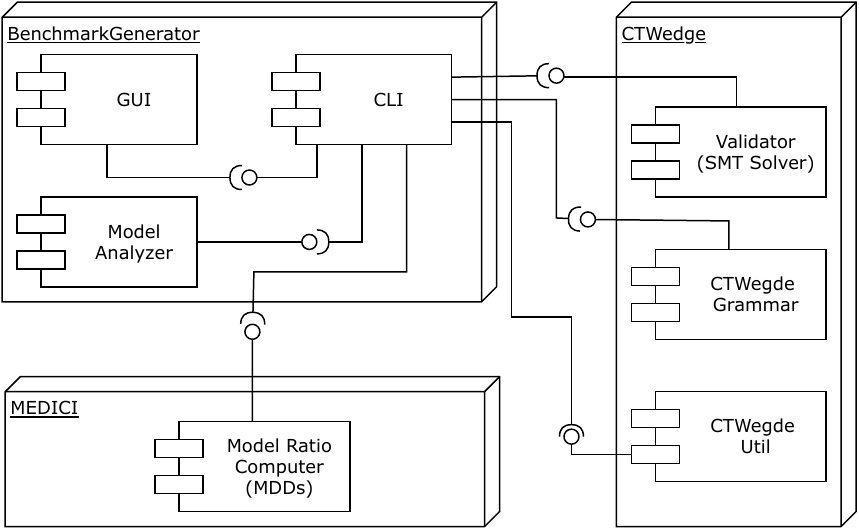}
	\caption{Software architecture of \tool}
	\label{fig:architettura}
\end{figure}

In this section, we describe the architecture we have designed for \tool, together with the strategies and approximations we used for computing relevant measures.
The tool architecture is reported in Fig.~\ref{fig:architettura}. \tool features a GUI and a CLI. The former aims at increasing the usability of the benchmark generator, but the business logic is completely implemented in the CLI.
The latter includes all the functionalities of the benchmark generator, such as the pure generation, the check for the existence of at least a test derivable from the generated IPM (see Sect.~\ref{sec:ex}), the computation of the tuple validity ratio (see Sect.~\ref{sec:tp}) and test validity ratio (see Sect.~\ref{sec:ts}).

The basic functionalities used by \tool are offered by the CTWedge environment~\cite{Gargantini2018}, including the CTWedge grammar definition, the utility functions (such as those generating the tuples, converting a CTWedge model in other formats, etc.), and the validation functionalities (exploited for checking the solvability of an IPM).
In the following, we describe in detail the role of each component of the architecture, by explaining the method we have implemented for checking the solvability of an IPM and computing relevant ratios.

\subsection{Existence of at least a test}\label{sec:ex}

When a benchmark is generated, it is important to check its solvability, i.e., the existence of at least a test case derivable from the IPM.
This check is done by the CTWedge \emph{validator} module in Fig.~\ref{fig:architettura}, which exploits an SMT Solver\footnote{We use the following SMT solver: \url{https://github.com/sosy-lab/java-smt}}, as presented in~\cite{Bombarda2021}. 
\ins{In particular, an SMT solver is a tool aiming to determine whether a mathematical formula is satisfiable or not, by using some modulo theories. In our case, the formula we want to check is a Boolean formula composed by the conjunction of all the constraints and defined on the Cartesian product of all the domains of the parameters of the IPM under analysis.}

The process to be followed for \chg{defining}{determining} if at least a test case can be derived from an IPM is very straightforward.
Each IPM generated by \tool is translated in its own SMT context, containing all the variables and constraints of the IPM. 
More in detail, the parameters of an IPM are translated into SMT variables depending on their type:
\begin{itemize}
	\item \emph{Booleans} are translated into SMT Boolean variables;
	\item \emph{Integer ranges} are translated into SMT integer variables. 
		Furthermore, since ranges in combinatorial models are limited between a lower and an upper bound, it is necessary to add to the context an additional constraint specifying these limits. For example, if a range is defined in the combinatorial model as $P1 : [-4 .. 3]$, in addition to the $P1$ integer variable, the following constraint is added: $P1 \geq -4$ AND $P1 \leq 3$;
	\item \emph{Enumeratives} are translated into SMT integer variables. 
		As for the normal integer ranges, when translating enumeratives, it is necessary to add to the SMT context a group of constraints limiting the values that can be taken by each enumerative.
		Furthermore, in this kind of transformation, it is necessary to use unambiguous numbers between parameters, in order to avoid different parameters assuming the same value.
		For example, if two enumeratives are defined in the combinatorial model as $P1 : \{A, B\}$ and $P2 : \{C, D\}$, the following is a valid mapping: $A \rightarrow 1$, $B \rightarrow 2$, $C \rightarrow 3$, and $D \rightarrow 4$. Moreover, for the parameter $P1$ the following constraint need to be added: $P1 \geq 1$ AND $P1 \leq 2$.
\end{itemize}

Additionally, all the other constraints of an IPM can be easily mapped to SMT formulas, exploiting the variables previously defined. 
Note that a combinatorial model may contain relational, mathematical, or comparison operators (between parameters or values) in general propositional formulas. 
All these aspects can be easily represented with operations between variables and values defined in an SMT context.

Then, if the context is SAT, it means that at least one test can be derived from the IPM and, thus, it can be accepted as benchmark. 

\subsection{Computation of the tuple validity ratio}\label{sec:tp}

To compute the tuple validity ratio $r_{tp}$ we exploit the same formalism presented in Sect.~\ref{sec:ex}, i.e., the  CTWedge \emph{validator} module in Fig.~\ref{fig:architettura}, similarly as done in~\cite{Bombarda2023}.
First, we build a complete SMT context $ctx$, containing all the parameters and constraints of the IPM, properly translated in SMT notation.
Then, we iterate over all the $t$-uples $tp_i \in TP$ and we check if adding $tp_i$ to $ctx$ makes the context still satisfiable.
In that case, it means that $tp_i$ is valid, otherwise it is not. 
By doing so, we compute the number of valid $t$-uples $v$ and, consequently, the tuple validity ratio $r_{tp}$ as follows:

\[
r_{tp}=\frac{v}{\#TP}
\]

\subsection{Computation of the test validity ratio}\label{sec:ts}

One of the desired characteristics of the benchmark models is the test validity ratio $r_{ts}$, introduced and defined in Sect.~\ref{sec:background}. Only for small models the calculation of $r_{ts}$ could be done by simply enumerating all the possible configurations and checking how many of them are valid.
For large models, we have devised two techniques, one that is precise, but it is not suitable for any model, while the other is approximate, but it can be used even when the model contains arithmetic constraints. 

\subsubsection{Using MDDs}

\begin{figure}
	\centering
	\begin{tikzpicture}[->,>=stealth',shorten >=1pt,auto,node distance=2cm,
		semithick]
		\tikzstyle{every state}=[text=black]
		
		\node[place] (a) {$a$};
		\node[place] (b) [right of=a] {$b$};
		\node[place] (c) [right of=b] {$c$};
		\node[place] (b1) [below of=b] {$b$};
		\node[place] (c1) [below of=c] {$c$};
		\node[place] (c2) [below of=c1] {$c$};
		\node[rectangle] (F) [right of=c2] {\textbf{F}};
		\node[rectangle] (T) [right of=c1] {\textbf{T}};
		
		\path
		(a) edge[above] node{$F$} (b)
		(b) edge[above] node{$*$} (c)
		(c) edge[bend left, right] node{$*$} (T)
		(a) edge[bend right, left] node{$T$} (b1)
		(b1) edge[above] node{$T$} (c1)
		(c1) edge[above] node{$*$} (T)
		(b1) edge[bend right, left] node{$F$} (c2)
		(c2) edge[above] node{$*$} (F);
		
	\end{tikzpicture}
	\caption{\ins{MDD structure for the combinatorial problem in Listing~\ref{lst:IPMForMDD}. With $*$ we mean all the possible values}}
	\label{fig:mdd}
\end{figure}
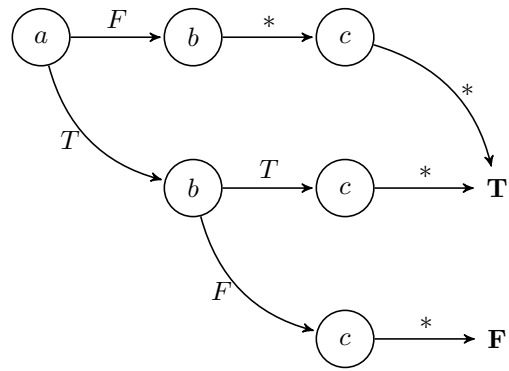

To count how many combinations are valid, we rely on a data structure, called Multi-Valued Decision Diagrams, on which the MEDICI~\cite{Gargantini2014} test generator (see Fig.~\ref{fig:architettura}) is based. 
\ins{Indeed, most combinatorial problems can be easily represented by using an MDD identifying valid combinations that comply with the constraints of the IPM under analysis.
Let's consider the IPM in Listing~\ref{lst:IPMForMDD}, which represents a combinatorial model with three parameters and a very simple constraint between $a$ and $b$.
With an MDD, as reported in Fig.~\ref{fig:mdd}, we can represent the validity of different parameter combinations.
By counting how many paths lead to the $T$ leaf, we can simply determine the number of valid tests without the need to generate each possible configuration and check if it is valid or not.}

\begin{lstlisting}[language=CTWedge, basicstyle=\sffamily\scriptsize, caption=\ins{Example of a simple IPM in CTWedge format},float, label=lst:IPMForMDD]
	Model example2
	
	Parameters:
	a : Boolean
	b : Boolean
	c : {V1, V2, V3}
	
	Constraints:
	# a => b #
\end{lstlisting}

\chg{After}{More in details, after} having generated an IPM $M$, we can execute MEDICI with the option \texttt{--donotgenerate}. 
In this way, MEDICI translates $M$ into its MDD representation, by starting with the definition of the nodes corresponding to the variables of the combinatorial model.
The cardinality of the MDD (i.e., the number of paths starting from the root node to the true leaf) is the number of all the possible tests $N$.
Then, we incrementally add all the constraints of $M$, and we compute again the cardinality of the MDD after all the constraints have been added.
This second cardinality corresponds to the number of valid tests $V$ for $M$ when the constraints are considered.
Thus, the test validity ratio is computed as follows:

\[
r_{ts}=\frac{V}{N}
\]

We emphasize that the cardinality of an MDD is not computed by \chg{enumeration}{enumerating all the possible assignments leading to the true leaf (although this would be possible~\cite{Toda2016})}, but recursively visiting the MDD and computing the final cardinality by sums and products of the cardinality of partial MDDs, thus the complexity of this operation is much lower than that of \chg{test}{path} enumeration.
However, although MDDs are very efficient in subset counting, not all combinatorial problems can be easily represented by an MDD. 
Indeed, as presented in~\cite{Bombarda2022a,Bombarda2023a}, MDDs allow users \chg{to only represent}{to represent in an optimized and memory-effective way only} combinatorial problems not containing arithmetical or comparison operations between parameters and values (e.g., $+$, $-$, $>$, $<$, etc.), or constraints comparing two different parameters (e.g., \lstinline[language=CTWedge]|PAR1 = PAR2|)).
\ins{Indeed, even if using MDDs would be technically feasible in those cases, we may likely have the problem of the combinatorial explosion of the number or complexity of constraints, thus leading to the impossibility of completely representing the combinatorial problem.}

\subsubsection{Using a Monte Carlo approach}

When the MDD-based technique presented before is not applicable, we can rely on one of the basic approximate set counting algorithms that are based on the classical Monte Carlo method. 
These methods can be applied because we have a finite set, containing all the possible tests, $U$ of known size $N$, and an efficient method for randomly choosing elements in $U$. 
We have also an efficient method to discover if a random test is valid or not (without using the solver, but simply by checking the truth value of each constraint when the assignments contained in the tests are set). 

To estimate the ratio $r_{ts}$, we can simply take a sequence of $n$ independent random tests by assigning a random value to each parameter in the model. 
Then we check if every test ${ts}_i$ is valid or not, and we assign to $x_i$ the value $0$ if the i-th test is not valid, otherwise, we assign to $x_i$ the value $1$. The total number of valid tests is $\sum_{1}^{n}x_i$. 

The Monte Carlo-based estimator for $r_{ts}$ that we indicate as $\tilde{r}_{ts}$  is simply:

\[
\tilde{r}_{ts}=\frac{\sum_{1}^{n}x_i}{n}
\]

It can be easily proved that this estimator is \emph{unbiased}, i.e., as the sample size $n$ increases the variance of the estimator decreases, improving the confidence of the estimation. 
\chg{If we could have $n=N$, the number of all the tests,}{If we could take all the possible $N$ tests and count how many of them are valid, then} we would get the right estimation.  
In most cases, we can only guarantee that the approximation is good enough if we take enough elements. 
In particular, the Zero-One Estimator Theorem~\cite{Karp1989a} gives us a lower bound for the number of elements to be considered in order to make a correct prediction with probability $p > 0$ and a maximum error of $\varepsilon > 0$, i.e.
\[
n \geq \dfrac{1}{r_{ts}}\cdot\dfrac{4 \cdot \ln{\dfrac{2}{1-p}}}{\varepsilon^2}
\]
If this requirement is satisfied, then, our prediction $\tilde{r}_{ts}$ is a correct approximation of the ratio $r_{ts}$ with probability:
\[
Pr[(1-\varepsilon) \cdot r_{ts} \leq \tilde{r}_{ts} \leq (1+\varepsilon) \cdot r_{ts}] \geq p
\]
%
%


When the user sets the desired ratio $r_{ts}$, we ask him/her to insert the desired probability $p$ and to set the acceptable error $\varepsilon$, so we can compute for every model $M$ the number $n$ of samples needed for making a prediction which is a correct approximation. Then, after having estimated $\tilde{r}_{ts}$ we check whether it is included in the range $(1-\varepsilon) \cdot r_{ts} \leq \tilde{r}_{ts} \leq (1+\varepsilon) \cdot r_{ts}$.
If the answer is yes, then we consider the model as satisfying the desired ratio $r_{ts}$. Otherwise, a new IPM has to be generated.

\smallskip
\begin{example}
	For a desired IPM $M$, the user asks for $r_{ts}=0.1$, $p=75 \%$, and $\varepsilon=0.1$. 
	The generator computes the number of required samples for making a correct approximation
	
	\[
		n \geq \dfrac{1}{r_{ts}}\cdot\dfrac{4 \cdot \ln{\dfrac{2}{1-p}}}{\varepsilon^2}=8,317.77
	\]
	
	Thus, the generator takes $8,318$ random tests, and let's assume that $825$ of them are valid, while $7,493$ are invalid. The estimated ratio is $\tilde{r}_{ts} = 0.099$, which is included in the range $\lbrack(1-\varepsilon) \cdot r_{ts}, (1+\varepsilon) \cdot r_{ts}\rbrack = \lbrack 0.09, 0.11 \rbrack$.
	Therefore, we can say that $M$ has ratio $r_{ts}=0.1$ with probability  $p \geq 75 \%$.
\end{example}

\section{Implementation}\label{sec:implementation}

\begin{figure*}[!tb]
	\centering
	\includegraphics[width=\linewidth]{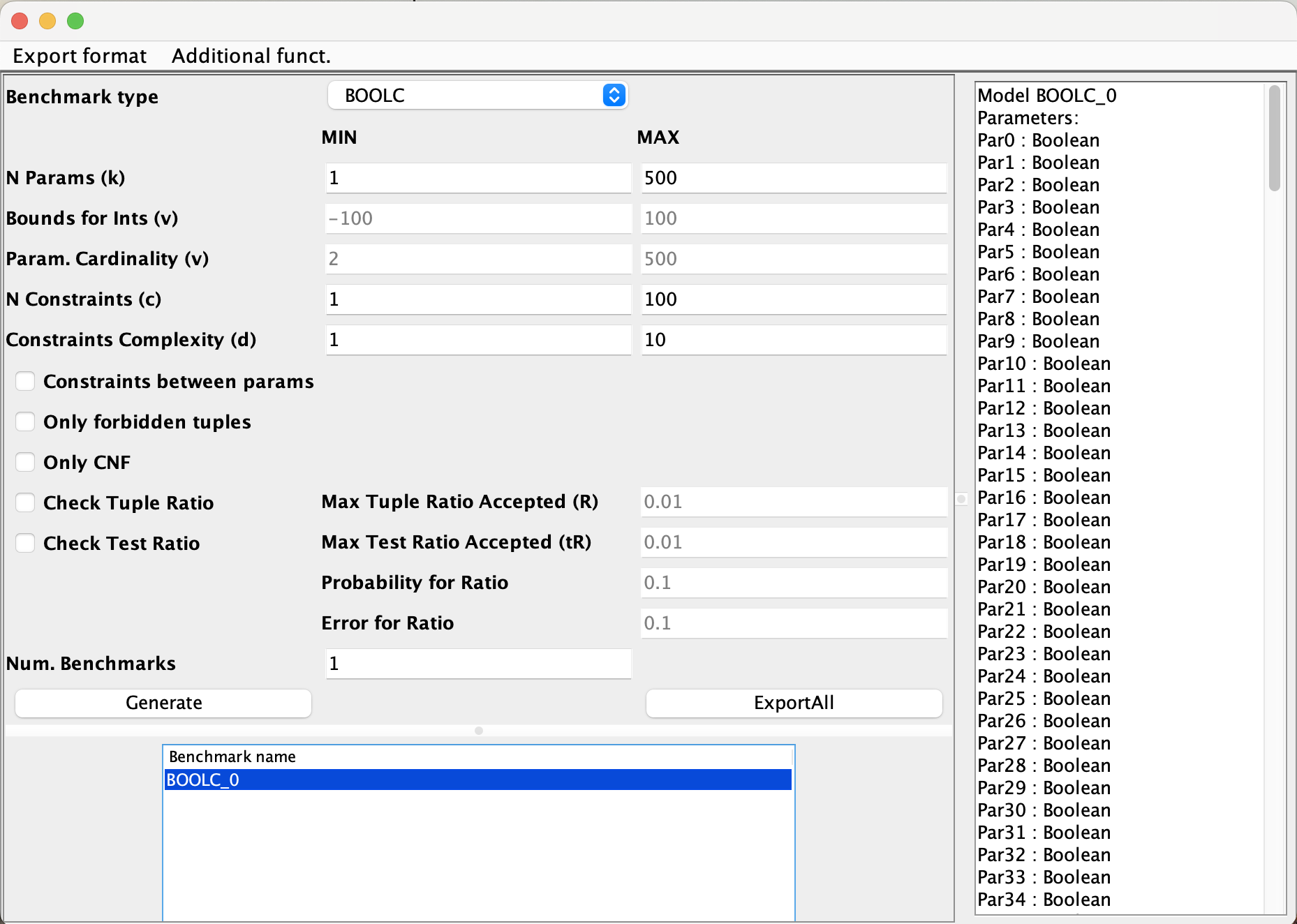}
	\caption{The \tool GUI}
	\label{fig:tool}
\end{figure*}

This section describes the implemented tool for generating benchmark IPMs, available as a command-line tool and with a GUI (see Fig.~\ref{fig:tool}).
In both versions, the generator allows the user to work in two different ways: 
\begin{itemize}
	\item The parameters of interest, depending on the chosen benchmark type, can be manually set;
	\item The parameters of interest, including the benchmark type, can be automatically set by giving a baseline model in CTWedge format, which is analyzed by \tool that extracts all the configuration parameter values \ins{(see Sect.~\ref{sec:modelanalyzer} for further details)}.
\end{itemize}

After having fixed the parameters of interest, the benchmarks are randomly generated by \tool.
To give an intuition on how the benchmark generator produces the models, in Algorithm~\ref{alg:NUMC}, we report the algorithm used for generating NUMC benchmarks. 
Note that the procedure is the same for the other benchmark categories, except for the type of constraints and parameters chosen.
\begin{algorithm*}[!tb]
	\caption{Algorithm for the generation of NUMC benchmarks}\label{alg:NUMC}
	\begin{algorithmic}[1]
		\Require $nBenchmarks$, the number of IPMs to be generated
		\Require $\langle kMin, kMax \rangle$, the min. and max. number of parameters for each IPM
		\Require $\langle lInt, uInt \rangle$, the lower and upper bounds for integer ranges
		\Require $\langle vMin, vMax \rangle$, the min. and max. cardinalities
		\Require $\langle cMin, cMax \rangle$, the min. and max. number of constraints for each IPM
		\Require $\langle dMin, dMax \rangle$, the min. and max. constraint complexities
		\Require $useCBtwP$, whether to use constraints between parameters
		\Require $FT$, whether to use only forbidden tuples in constraints		
		\Require $CNF$, whether to use only constraints in CNF
		\Require $r_{tp}$, the max. tuple validity ratio
		\Require $useTupleRatio$, whether to consider $r_{tp}$ during IPMs generation
		\Require $\langle r_{ts}, p, \varepsilon \rangle$, the max. test validity ratio, the probability, and the maximum error for the ratio
		\Require $useTestRatio$, whether to consider $r_{ts}$ during IPMs generation
		\Ensure $modelsList$, the list of the generated benchmarks 
		\StateX
		\StateX \LeftComment{-1}{Initially, no models have been generated}
		\State $modelsList \gets \emptyset$; $nB \gets 0$; 
		\While{$nB < nBenchmarks$}
		
		\For{$nAttempts \gets 1$ to $10$}                    
		
		\StateX \LeftComment{1}{Randomly define parameters}
		\State $nParams \gets$ \Call{randomBetween}{$kMin$,$kMax$} \label{alg:randParam}
		\State $pList \gets$ \Call{defineParams}{$nParams$, $lInt$, $uInt$, $vMin$, $vMax$} \label{alg:randParamGen}
		\StateX \LeftComment{1}{Randomly define constraints}
		\State $nC$ $\gets$ \Call{randomBetween}{$cMin$,$cMax$} \label{alg:randCnstr}
		\State $cList$ $\gets$ \Call{defineCnstr}{$pList$, $nC$,  $dMin$, $dMax$, $useCBtwP$, $FT$, $CNF$}\label{alg:randCnstrGen}
		\StateX \LeftComment{1}{Check that the generated IPM complies with the requirements}
		\State $model$ $\gets$ \Call{buildModel}{$pList$, $cList$} \label{alg:model}
			\If {$model$.isSolvable()} \label{alg:solvable} 
			\If {\Not $useTupleRatio$ \Or $model$.getTupleValidityRatio() $< r_{tp}$} \label{alg:rtp}
				\If {\Not $useTestRatio$ \Or $model$.getTestValidityRatio($p$, $\varepsilon$) $< r_{ts}$}  \label{alg:rts}
					\State $modelsList$.add($model$)
					\State $nB \gets nB + 1$; $nAttempts \gets 0$
					\State \textbf{break}
				\EndIf
			\EndIf 	\EndIf	
		\EndFor
		
	\EndWhile
\end{algorithmic}
\end{algorithm*}

The algorithm aims at producing $nBenchmarks$ IPMs with the desired characteristics.
For each benchmark, initially, the tool extracts a random number of parameters (line~\ref{alg:randParam}) with bounds $kMin$ and $kMax$.
Then, the set of the parameters to be included in the IPM is generated by the function \textsf{defineParams}, which randomly extracts the types and values for each parameter (line~\ref{alg:randParamGen}). 
The same approach is followed for constraints definition (lines~\ref{alg:randCnstr} and~\ref{alg:randCnstrGen}).
In Sect.~\ref{sec:param} we will explain in detail the algorithm defining the parameters, and in Sect.~\ref{sec:constraints} that defining the constraints.
In this way, \tool produces a single IPM (line~\ref{alg:model}) which now needs to be checked to see whether it is solvable (line~\ref{alg:solvable}) and, if the tuple validity ratio and/or the test validity ratio have to be met, possibly has the required ratios (lines~\ref{alg:rtp} and~\ref{alg:rts}).
In that case, the model is added to the $modelsList$, otherwise, a new model is generated.

This process can last for a long time, especially if some check on the ratio is required. For this reason, we set a maximum number of $nAttempts$ of 10 trials for the single IPM. 
Note that different approaches may be used, especially when considering the ratio of IPMs, for producing only benchmarks complying with the requirements, such as adding one constraint per time and building incrementally the model.
However, this may cause to be stuck in models where no constraint making the model solvable or complying with the ratio required can be added.
As a future work, we may investigate this approach in order to solve its limitations (e.g., by using a backtracking strategy) and to avoid completely throwing away the generated IPM every time it is not compliant with the characteristics requested by the user.

\subsection{Parameters definition}\label{sec:param}

For every benchmark IPM, after having fixed the number of parameters, \tool defines the type and values for each of them randomly.
The type, and consequently the values, of each parameter depends on the requested type of benchmarks (e.g., for UNIFORM\_BOOLEAN only Boolean parameters are chosen, for MCA and MCAC, the type of each parameter is chosen between Boolean and enumeratives, while for NUMC also integer ranges are considered).
As previously done for the general algorithm, we here give an explanation of the parameters' definition algorithm for NUMC IPMs in Alg.~\ref{alg:ParameterDefinition}, but for the other categories, the procedure is the same, except that fewer types of parameters are used.
\begin{algorithm*}[!tb]
	\caption{Algorithm for the definition of parameters in the case of NUMC benchmarks}\label{alg:ParameterDefinition}
	\begin{algorithmic}[1]
		\Require $nParams$, the number of parameters
		\Require $\langle lInt, uInt \rangle$, the lower and upper bounds for integer ranges
		\Require $\langle vMin, vMax \rangle$, the min. and max. cardinalities
		\Ensure $paramsList$, the list of the random parameters 
		\StateX
		\Function{defineParameters}{$nParams$, $lInt$, $uInt$, $vMin$, $vMax$}
		\StateX \LeftComment{0}{Initially, no parameters have been defined}
		\State $paramsList \gets \emptyset$; $nP \gets 0$; 
		\While{$nP < nParams$}
		\StateX \LeftComment{1}{Extract the type of the current parameter}
		\State $pType \gets$ \Call{chooseRandom}{Boolean, Enumerative, Range} \label{alg:partype}
		\If {$pType = $ Boolean} \label{alg:bool}
			\StateX \LeftComment{2}{Boolean variables require only to set their name}
			\State $param \gets$ \Call{createNewBoolean}{``PAR'' + nP}
		\ElsIf{$pType = $ Enumerative} \label{alg:enum}
			\StateX \LeftComment{2}{Enumeratives require to set all possible values, which are random as well}
			\State $nValues \gets$ \Call{randomBetween}{$vMin$, $vMax$}
			\State $param \gets$ \Call{createNewEnum}{``PAR'' + nP}
			\State $i \gets 0$
			\For{$i < nValues$}
				\State $param$.values.add(``PAR'' + nP + ``\_'' + $i$)
				\State $i \gets i + 1$
			\EndFor
		\Else \label{alg:ranges}
			\StateX \LeftComment{2}{For integer ranges, we need to set the lower and upper bound}
			\StateX \LeftComment{2}{but the cardinality must be considered as well}
			\State $param \gets$ \Call{createNewRange}{``PAR'' + nP}
			\State $\langle l, u \rangle \gets$ \Call{randomRange}{$lInt$, $uInt$, $vMin$, $vMax$}
			\State $param$.setRange($l$, $u$)
		\EndIf
		\State $paramsList$.add($param$)
		\State $nP \gets nP + 1$
		\EndWhile
		\State \Return $paramsList$
		\EndFunction	
	\end{algorithmic}
\end{algorithm*}

In general, for each parameter, at line~\ref{alg:partype}, the algorithm randomly defines the parameter type (among Booleans, enumeratives, and integer ranges).
If the parameter is Boolean (line~\ref{alg:bool}) no additional setting is required. 
On the other hand, if an enumerative or range has to be created, additional information has to be set.
In the former case (line~\ref{alg:enum}), the number of values is randomly set within the bounds given by $vMin$ and $vMax$.
In the latter case (line~\ref{alg:ranges}), the bounds of the range have to be set by the function \textsf{randomRange}. 
At this stage, \tool considers both the integer bounds $lInt$ and $uInt$, but observes the cardinality bounds ($vMin$ and $vMax$) as well.

The described process is repeated for the number of parameters required and, then, at the end, a full $paramList$ is produced, containing parameters with different types and values.

\subsection{Constraints definition}\label{sec:constraints}

After having set the parameters of the IPM, it is necessary to add (whether applicable), the constraints.
As for parameters, the constraints are randomly defined, both in terms of number and complexity.
Instead, unlike the parameters, the constraints are all composed in a very similar way, regardless of the benchmark type. 
For NUMC IPMs, relational and mathematical operations are possible as well.

\begin{algorithm*}[!tb]
	\caption{Algorithm for the definition of constraints in the case of NUMC benchmarks}\label{alg:NUMCconstraints}
	\begin{algorithmic}[1]
		\Require $pList$, the list of parameters
		\Require $nCnstr$, the number of constraints to be generated
		\Require $\langle dMin, dMax \rangle$, the min. and max. constraint complexities
		\Require $CBtwP$, whether to use constraints between parameters
		\Require $FT$, whether to use only forbidden tuples in constraints
		\Require $CNF$, whether to use only constraints in CNF
		\Ensure $cList$, the list of the constraints generated 
		\StateX
		\Function{defineCnstr}{$pList$, $nCnstr$,  $dMin$, $dMax$, $CBtwP$, $FT$, $CNF$}
		\StateX \LeftComment{0}{Initially, no constraints have been defined}
		\State $cList \gets \emptyset$; $nC \gets 0$; 
		\While{$nC < nCnstr$}
			\StateX \LeftComment{1}{Randomly define the complexity}
			\State $compl \gets$ \Call{randomBetween}{$dMin$, $dMax$} \label{alg:complexity}
			\StateX \LeftComment{1}{Generate the constraint}
			\State $cList$.add(\Call{generateCnstr}{$pList$, $compl$, $CBtwP$, $FT$, $CNF$}
			\State $nC \gets nC + 1$
		\EndWhile
		\State \Return $cList$
		\EndFunction
		\StateX
		\Function{generateCnstr}{$pList$, $compl$, $CBtwP$, $FT$, $CNF$}
			\If {$compl > 1$} 
				\StateX \LeftComment{1}{Recursively define the constraint}
				\State $cType \gets $ \Call{chooseRandom}{AND, OR, IMPL, DBLIMPL} \label{alg:randomType}
				\State $c \gets$ \Call{createConstraintByType}{$cType$}
				\StateX \LeftComment{1}{Set the left and right part}
				\State $c$.setLeft(\Call{generateCnstr}{$pList$, $(compl-1)/2$, $CBtwP$, $FT$, $CNF$}) \label{alg:left}
				\State $c$.setRight(\Call{generateCnstr}{$pList$, $(compl-1)/2$, $CBtwP$, $FT$, $CNF$}) \label{alg:right}
			\Else
				\StateX \LeftComment{1}{Define an atomic constraint}
				\State $c \gets $ \Call{createAtomicConstraint}{$pList$, $CBtwP$, $FT$, $CNF$} \label{alg:atomic}
			\EndIf
			\State \Return $c$
		\EndFunction
	\end{algorithmic}
\end{algorithm*}

The constraints definition process is based on Alg.~\ref{alg:NUMCconstraints}. 
After having defined the number of constraints, as shown in Alg.~\ref{alg:NUMC}, the algorithm randomly chooses the complexity of every single constraint (line~\ref{alg:complexity}).
Then, the composition of the constraint is performed by the \textsf{generateConstraint} recursive function.
It is designed for composing the constraint as an AND, OR, implication, or double implication of atomic constraints (line~\ref{alg:randomType}).
This process is recursively repeated while the remaining complexity is greater than $1$ and populates, for each constraint, the left and the right part (line~\ref{alg:left} and~\ref{alg:right}).
Then, when the complexity reaches the value $1$, a single atomic constraint is created (line~\ref{alg:atomic}), in the form of $PAR = val$ or $PAR_X = PAR_y$ (or $\neq$, $>$, $\geq$, $<$, $\leq$, depending on the type of the IPM being generated).
Note that the decision on the operator to be used in the atomic predicate, as well as the decision on whether to compare two parameters or a parameter and a value is randomly made by the benchmark generator thanks to the function \textsf{createAtomicConstraint}.

\subsection{\ins{Model analyzer}}\label{sec:modelanalyzer}

\ins{In this section, we analyze the \emph{Model analyzer} component, which is used by \tool for automatically extracting the configuration depending on an already available CTWedge IPM $M$.}

\ins{For what concerns the benchmark type (as reported in Tab.~\ref{tab:tracks}), first, \tool looks for the constraints in $M$.
If no constraint is found, then, the decision on the benchmark type is taken depending on the type of parameters.
When all parameters are Booleans, the model is considered as an \texttt{UNIFORM\_BOOLEAN} instance; if all parameters are all with the same size, the model is considered as an \texttt{UNIFORM\_ALL} instance, while in all the other cases it is an \texttt{MCA} instance.
On the other hand, if constraints are present in $M$, then the benchmark type is within \texttt{BOOLC}, \texttt{MCAC}, or \texttt{NUMC}.
The first category is chosen when all parameters are Booleans; the second is assigned when not all parameters are Boolean but no integer ranges are available in $M$, while the last benchmark type is chosen in all the other cases.}

\ins{Depending on the benchmark type, identified by the Model analyzer, following actions are taken by \tool.
The number of parameters (both minimum and maximum - $k$ in the \tool GUI) are automatically set by counting the parameters in $M$, as well as done for the constraints ($c$ in the \tool GUI).
The minimum and maximum cardinality for the parameters ($v$ in the \tool GUI), or the bounds for integer ranges, are computed by enumerating all the parameters in $M$ and identifying the one with the lowest and the one with the highest cardinality. 
Regarding the minimum and maximum constraints complexity ($v$ in the \tool GUI), they are computed by applying Definition~\ref{def:complexity} to all the constraints in an iterative way, in order to find the lowest and highest values. More specifically, the Model analyzer, extracts the complexity from each single constraint by recursively visiting it and identifying the number of binary logical operators or connectors.}

\ins{The Model analyzer can also extract from $M$ the type of the constraints, i.e., if all of them are expressed as forbidden tuples or in CNF. This analysis is done by iteratively visiting all the constraints and exploiting the modelanalyzer utility in the CTWedge framework~\cite{Bombarda2021}.}

\ins{Finally, the \emph{tuple validity ratio} and the \emph{test validity ratio} are extracted by $M$ by applying the same strategies presented in Sec.~\ref{sec:tp} and Sect.~\ref{sec:ts}, and choosing the most suitable approach depending on the benchmark type, i.e., on the type of the parameters and constraints in $M$.}

\subsection{\ins{\tool usage}}

\ins{In this section, we delve into the workflow and usage of \tool and its GUI for generating a set of benchmarks. Additional instruction for the CLI version of \tool are available at \url{https://github.com/fmselab/CIT_Benchmark_Generator/tree/main/BenchmarkGenerator}.} 

\ins{First, the user needs to set the \emph{Benchmark type}, by choosing one of those proposed by \tool. 
In this way, the configuration parameters of interest in the left column (as shown in Fig.~\ref{fig:fm} and explained in Sect.\ref{sec:requirements}) are automatically enabled, filled with default values, and can be set by the user.
After having set all the parameters, the \emph{Generate} button allows for generating the benchmark IPMs complying with the chosen configuration parameters.}

\ins{When the generation process terminates, the names of the IPMs are shown in the list in the lower part of the left column of \tool, and the full model is shown when the user clicks on one of them.
The exporting process is very straightforward: first, the formats of interest are set through the \emph{Export format} button in the menu bar; then the IPMs are exported in the chosen formats when the user clicks on the \emph{ExportAll} button.}

\ins{If a baseline IPM is available and the user wants to generate new IPMs with the same characteristics, the model analyzer component introduced in Sect.~\ref{sec:modelanalyzer} can be triggered by clicking on the \emph{Set baseline IPM} button under the \emph{Additional funct.} menu in the menu bar.}

\ins{Finally, the use of a domain specific dictionary\footnote{Examples of dictionaries are available at: \url{https://github.com/fmselab/CIT_Benchmark_Generator/tree/main/BenchmarkGenerator/dictionaries}} is allowed by the \emph{Set dictionary} button under the \emph{Additional funct.} menu in the menu bar. 
When a dictionary is set, the name and values of the parameters are chosen among those provided in the dictionary, if available. Otherwise, the regular naming strategy is adopted.}

\section{Validation}\label{sec:validation}

In this section, we report how we have validated \tool by testing its functionalities and \ins{ensured that generated benchmarks reflect real-world software systems' characteristics} by showing that the majority of the benchmarks available in the literature can be generated by our tool. 

\begin{table*}[tb!]
	\centering
	\caption{The test suite derived from the feature model of the requirements}
	\label{tab:testSuite}
	\setlength{\tabcolsep}{1.5pt}
	\renewcommand{\arraystretch}{1.2}
	\begin{tabular}{l|ccccccccccccccccc}
		\toprule
		\textbf{Param} & \rotatebox[origin=c]{90}{\textbf{$ts_1$}} & \rotatebox[origin=c]{90}{\textbf{$ts_2$}} & \rotatebox[origin=c]{90}{\textbf{$ts_3$}} & \rotatebox[origin=c]{90}{\textbf{$ts_4$}} & \rotatebox[origin=c]{90}{\textbf{$ts_5$}} & \rotatebox[origin=c]{90}{\textbf{$ts_6$}} & \rotatebox[origin=c]{90}{\textbf{$ts_7$}} & \rotatebox[origin=c]{90}{\textbf{$ts_8$}} & \rotatebox[origin=c]{90}{\textbf{$ts_9$}} & \rotatebox[origin=c]{90}{\textbf{$ts_{10}$}} & \rotatebox[origin=c]{90}{\textbf{$ts_{11}$}} & \rotatebox[origin=c]{90}{\textbf{$ts_{12}$}} & \rotatebox[origin=c]{90}{\textbf{$ts_{13}$}} & \rotatebox[origin=c]{90}{\textbf{$ts_{14}$}} & \rotatebox[origin=c]{90}{\textbf{$ts_{15}$}} & \rotatebox[origin=c]{90}{\textbf{$ts_{16}$}} & \rotatebox[origin=c]{90}{\textbf{$ts_{17}$}} \\
		\midrule
		\textbf{\#B} & X & X & X & X & X & X & X & X & X & X & X & X & X & X & X & X & X \\
		\textbf{Type} & NC & NC & NC & MC & MC & MC & BC & BC & BC & M & UA & UB & BC & UA & M & UB & MC \\
		\textbf{Ratio} & - & X & - & X & - & X & - & X & X & - & - & - & X & - & - & - & X \\
		\textbf{$r_{tp}$} & - & X & - & X & - & X & - & X & - & - & - & - & X & - & - & - & - \\
		\textbf{$r_{ts}$} & - & X & - & - & - & X & - & X & X & - & - & - & X & - & - & - & X \\
		\textbf{Int.Bounds} & X & X & X & - & - & - & - & - & - & - & - & - & - & - & - & - & - \\
		\textbf{Card.} & X & X & X & X & X & X & - & - & - & X & X & - & - & X & X & - & X \\
		\textbf{\#P} & X & X & X & X & X & X & X & X & X & X & X & X & X & X & X & X & X \\
		\textbf{CnstrConf} & X & X & X & X & X & X & X & X & X & - & - & - & X & - & - & - & X \\
		\textbf{BtwParam} & - & X & X & - & X & - & X & - & X & - & - & - & X & - & - & - & - \\
		\textbf{Complx.} & X & X & X & X & X & X & X & X & X & - & - & - & X & - & - & - & X \\
		\textbf{\#C} & X & X & X & X & X & X & X & X & X & - & - & - & X & - & - & - & X \\
		\textbf{CnstrForm} & G & C & F & G & C & F & G & C & F & - & - & - & G & - & - & - & F \\
		\textbf{Params.} & X & X & X & X & X & X & X & X & X & X & X & X & X & X & X & X & X \\
		\textbf{Ranges} & X & X & X & - & - & - & - & - & - & - & - & - & - & - & - & - & - \\
		\textbf{Enums.} & X & X & X & X & X & X & - & - & - & X & X & - & - & X & X & - & X \\
		\textbf{Booleans} & X & X & X & X & X & X & X & X & X & X & X & X & X & X & X & X & X \\
		\textbf{Cnstr.} & X & X & X & X & X & X & X & X & X & - & - & - & X & - & - & - & X \\
		\textbf{Arithmetic} & X & X & X & - & - & - & - & - & - & - & - & - & - & - & - & - & - \\
		\textbf{Ex.Format} & - & X & X & - & X & - & X & - & - & X & - & X & X & X & - & - & X \\
		\textbf{ACTS} & - & X & X & - & X & - & - & - & - & X & - & X & X & X & - & - & - \\
		\textbf{CTWedge} & - & X & - & - & X & - & X & - & - & X & - & X & - & X & - & - & X \\
		\textbf{PICT} & - & X & X & - & - & - & X & - & - & X & - & X & X & X & - & - & X \\ \midrule
		\textbf{\textit{Outcome}} & \checkmark & \checkmark & \checkmark & \checkmark & \checkmark & \checkmark & \checkmark & \checkmark & \checkmark & \checkmark & \checkmark & \checkmark & \checkmark & \checkmark & \checkmark & \checkmark & \checkmark  \\
		\bottomrule
	\end{tabular}%
\end{table*}

\subsection{CIT for validation}

In order to validate and test \tool, we have applied a dogfooding technique: we derive from the feature model in Fig.~\ref{fig:fm}, describing the requirements of our tool, a combinatorial test suite with strength $t=2$.
The test suite has been generated, after having automatically translated the feature model in a CTWedge model, using ACTS, and, with only $17$ tests it allowed us to effectively test \tool.
The test cases are reported in Tab.~\ref{tab:testSuite}, where $\#B$ indicates the number of benchmarks, the $Type$ is expressed with the abbreviations introduced in Tab.~\ref{tab:tracks}, $\#P$ represents the number of parameters, $\#C$ the number of constraints, and the ConstraintForm is $C$ if constraints need to be in CNF, $G$ if the general form is required, or $F$ if forbidden tuples are used.
Note that abstract features (those in light blue in the feature model in Fig.~\ref{fig:fm}) are not reported in the test suite, since they are not actual features of the generators, but they are only used for grouping other features.

\begin{table*}[tb]
	\centering
	\caption{Summary of the values set for each non-boolean feature during test execution}
	\label{tab:values}
	\begin{tabular}{l|cccccccc}
		\toprule
		& \textbf{\#B} & \textbf{$r_{tp}$} & \textbf{$r_{ts}$} & \textbf{Int.Bounds} & \textbf{Card.} & \textbf{\#P} & \textbf{Complx.} & \textbf{\#C} \\ \midrule
		\textbf{Min} & - & - & - & -50 & 2 & 2 & 1 & 1 \\
		\textbf{Max} & - & - & - & 50 & 30 & 30 & 15 & 20 \\
		\textbf{$p$} & - & - & 75.0 \% & - & - & - & - & - \\
		\textbf{$\varepsilon$} & - & - & 0.1 & - & - & - & - & -\\
		\textbf{Value} & 10 & 0.1 & 0.1 & - & - & - & - & -\\
		\bottomrule
	\end{tabular}%
\end{table*}

Some of the parameters that can be selected or unselected in the generated test suite actually correspond to many parameters that have to be set during test execution (e.g., the number of parameters $\#P$ requires to set the maximum and minimum number).
Therefore, in Tab.~\ref{tab:values} we report the values we set in each test case for each non-boolean feature, but we emphasize that these values are reported only for completeness and replicability of the tests, and the same results would be obtained with every other values. 
Note that we decided to use $\#B = 10$ in order to have multiple examples to check for every test case, considering that models are generated randomly by the benchmark generator.
Then, for each test case $ts_i$, after having set all the configuration parameters, we generate the benchmarks and check that every generated IPM conforms to its expected properties, in terms of parameters, constraints, ratio, and complexity.

The code executing the tests is available online in \tool's official repository \url{https://github.com/fmselab/CIT_Benchmark_Generator}, while the outcome of each test execution is reported in the last row of Tab.~\ref{tab:testSuite}.

\subsection{External validation}\label{sec:external}

Ensuring the similarity between artificially generated benchmarks and real models is of utmost importance when evaluating generators. 
This is crucial to avoid bias in the evaluation process, as models that do not accurately represent real systems can introduce distortions in the assessment of generator performance and correctness.
For this reason, in this section, we show that a significant number of models taken from the literature can be obtained by at least one configuration of our CIT benchmark generation.

\begin{table*}[tb!]
	\centering
	\caption{Summary values for the IPMs taken from the literature. In columns where two values are reported, they represent the lower and the upper bounds. FT reports the number of models in that category having forbidden tuples, while CF those with constraints in CNF}
	\label{tab:externalValidation}
	\setlength{\tabcolsep}{1.6pt}
	\begin{tabular}{ccccccccccccc}
		\toprule
		\textbf{Src} & \textbf{\#Ms} & \textbf{Type} & \textbf{\#} & \textbf{\#P} & \textbf{Int bnd.} & \textbf{Card.} & \textbf{\#C} & \textbf{Comp.} & \textbf{FT} & \textbf{CF} & \textbf{$r_{tp}$} & \textbf{$r_{ts}$} \\ \midrule
		\cite{petke2015practical} & 7 & BC & 1 & 10 & -- & -- & 1 & 5 & 0 & 1 & 0.99 & 0.75 \\
		& & MC & 6 & 7-14 & -- & 2-10 & 6-83 & 1-38 & 0 & 6 & 0.75-0.92 & 0.002-0.250\\ \midrule
		\cite{Jin2020}* & 11 & BC & 7 & 65-1,639 & -- & -- & 108-4,664 & 1-100 & 0 & 0 & 0.70-0.93 & 0.000-0.000 \\ 
		&     & MC & 4 & 72-6,295 & -- & 2-27 & 94-9,842 & 1-352 & 0 & 0 & 0.63-0.82 & 0.000-0.000 \\ \midrule
		\cite{Segall2011} & 18 & BC & 1 & 5 & -- & -- & 7 & 1-6 & 0 & 1 & 0.90 & 0.250 \\
		& & MC & 17 & 4-35 & -- & 2-13 & 3-388 & 1-8 & 13 & 17 & 0.75-1.00 & $10^{-5}$-0.654 \\ \midrule
		\cite{Garvin2010} & 35 & MC & 35 & 30-199 & -- & 2-6 & 5-49 & 1-9 & 0 & 35 & 0.80-0.99 & $10^{-5}$-0.324 \\ \midrule
		\cite{PICTRepo} & 28 & M & 7 & 6-18 & -- & 1-7 & -- & -- & -- & -- & -- & -- \\ 
		&     & BC & 1 & 7 & -- & -- & 2 & 1 & 0 & 0 & 0.98 & 0.625 \\ 
		&     & MC & 20 & 2-33 & -- & 1-11 & 1-36 & 1-9 & 0 & 0 & 0.80-0.99 & $10^{-7}$-0.813 \\  \midrule
		\cite{TzorefBrill2018} & 112 & M & 15 & 3-61 & -- & 1-500 & -- & -- & -- & -- & -- & -- \\ 
		&     & BC & 1 & 23 & -- & -- & 19 & 1-4 & 0 & 0 & 0.94 & 0.024 \\ 
		&     & MC & 93 & 4-118 & -- & 1-166 & 1-381 & 1-252 & 4 & 4 & 0.08-0.99 & $10^{-13}$-1.000 \\ 
		&     & NC & 2 & 8-8 & 0-3 & 2-4 & 7-11 & 1-4 & 0 & 0 & 0.35-0.91 & 0.088-1.000 \\ \midrule
		\cite{Johansen2011}* & 16 & BC & 8 & 28-1,397 & -- & -- & 34-3,633 & 1-13 & 0 & 0 & 0.61-0.93 & $10^{-12}$-0.001\\ 
		&     & MC & 8 & 7-6,295 & -- & 2-27 & 7-9,842 & 1-352 & 0 & 0 &  0.47-0.82 & $10^{-13}$-0.103 \\ \midrule
		\cite{CTCompetition2022Repo} & 300 & UB & 53 & 2-20 & -- & -- & -- & -- & -- & -- & -- & -- \\ 
		&     & UA & 50 & 2-20 & -- & 2-20 & -- & -- & -- & -- & -- & -- \\ 
		&     & M & 47 & 2-20 & -- & 2-50 & -- & -- & -- & -- & --  & --\\ 
		&     & BC & 54 & 2-20 & -- & -- & 1-23 & 1-15 & 0 & 0 & 0.25-1.00 & $10^{-5}$-0.937 \\ 
		&     & MC & 60 & 2-19 & -- & 2-50 & 1-38 & 1-14 & 0 & 4 & 0.01-1.00 &  $10^{-16}$-1.000 \\ 
		&     & NC & 36 & 2-17 & (-99)-100 & 1-199 & 1-13 & 1-14 & 1 & 1 & 0.01-1.00 & 0.001-1.000 \\ \midrule
		\cite{CTCompetition2023Repo} & 240 & UB & 15 & 7-29 & -- & -- & -- & -- & -- & -- & -- & -- \\ 
		&     & UA & 15 & 7-25 & -- & 2-15 & -- & -- & -- & -- & -- & --\\ 
		&     & M & 30 & 7-30 & -- & 1-15 & -- & -- & -- & -- & -- & --\\ 
		&     & BC & 36 & 6-43 & -- & -- & 1-46 & 1-14 & 0 & 1 & 0.33-1.00 & $10^{-9}$-0.906 \\ 
		&     & MC & 114 & 4-199 & -- & 1-15 & 1-37 & 1-15 & 0 & 44 & 0.02-1.00 & $10^{-12}$-1.000 \\ 
		&     & NC & 30 & 6-30 & (-100)-111 & 1-16 & 1-24 & 1-14 & 0 & 0 & 0.06-1.00 & 0.001-0.830\\ 
		\bottomrule
	\end{tabular}%
\end{table*}

Tab.~\ref{tab:externalValidation}\footnote{For the models in the NUMC category, $r_{ts}$ is an estimation computed with the Monte Carlo-Based approach, as explained in Sect.~\ref{sec:ts}, with $n=1000$. For some models (those with the *) it was not possible to compute both $r_{ts}$ and $r_{tp}$ because of their high complexity~\cite{Thm2020}.} shows the 767 models we have considered and the characteristics extracted from them by the \emph{modelanalyzer} part of our CIT benchmark generator, i.e., the part meant to extract the configuration from a given IPM where the generation from a baseline model is chosen (see Sect.~\ref{sec:implementation}).
All models and data extracted from their analysis are available at \url{https://github.com/fmselab/CIT\_Benchmark\_Generator/blob/main/BenckmarkGenerator/external\_validation}.

Data reported in Tab.~\ref{tab:externalValidation} show that in all the considered cases, we have been able to classify the models from the literature in the categories handled by \tool, and all the categories we can generate with \tool have been found in the literature.
The only limit we found is dealing with very complex models having thousands of parameters and constraints (derived from Software Product Lines and not natively representing IPMs, though), for which computing the ratio is not feasible in an exact way.
Considering the data obtained by analyzing the IPMs available in the literature, we can conclude that by setting \tool in the same way as in those benchmarks, we can obtain plausible models with the same features as real-world IPMs. 

\section{Related work}\label{sec:related}

Benchmarking combinatorial test generators is of paramount importance since it allows both for assessing the correctness of the tools (i.e., their ability to produce valid and complete test suites, covering all the desired $t$-way interactions) and for evaluating their performance.
Several works have been presented in the past, trying to evaluate test generators and identifying those having the best performance, both in terms of generation time and test suite size.
For example, in~\cite{Bombarda2021}, the authors presented a benchmarking environment, based on CTWegde~\cite{Gargantini2018} which allows the comparison between test generators that can be easily included by extending some selected Eclipse extension points. In that work, the authors compared some of the most well-known generators (ACTS~\cite{Yu2013}, MEDICI~\cite{Gargantini2014}, CAgen~\cite{Wagner2020}, PICT~\cite{PICTRepo}, and CASA~\cite{Garvin2009}), but only on a limited set of 196 IPMs taken from the literature.

In this paper, instead, we focus more on generating benchmark models and not on their execution for comparing test generators.
Indeed, finding real IPMs is not so easy in the literature, since many of those used in research works are not distributed due to IP limitations. 
Some analyses, when real highly configurable system models are needed, have been conducted by deriving combinatorial models from software product lines, such as in~\cite{Johansen2011} and~\cite{Jin2020}.
This is not always the optimal approach, since the translation of an SPL into an IPM requires some assumption (such as the way in which alternative groups are translated, or the way in which abstract or hidden features are treated) that may vary the complexity of the generated IPMs.

This is the reason why we focus on benchmark generation.
This problem is not completely new and it is tackled also by other works.
For example, in~\cite{Younes2005}, the authors proposed a method for generating benchmarks, with known solutions, that does not suffer the usual limitations on the problem size or the sequence length, since it does not require the re-optimization phase.
This approach is different w.r.t. that we use in this paper since we do not require any solution to be known and, thus, we can generalize better test models.
Moreover, in~\cite{Ansotegui2023}, the authors propose a generator for benchmark IPMs, but only a limited set of features is addressed. For example, when considering constraints, only models containing Boolean parameters can be generated, while the tool presented in our paper supports also enumeratives and integer ranges.

Benchmark generation is a common approach for comparing different methods, techniques, and tools~\cite{Hasselbring2021}. It has been widely adopted especially in the context of competitions but not only. For instance, in \cite{Derks2023} the authors introduce \texttt {vpbench}, which simulates the evolution of a variant-rich system. The tool generates an evolution together with metadata that explains it - like in our case we generate a benchmark together with its type. 
In~\cite{Ferrer2011}, an automatic benchmark generator of java programs is presented. As done in our work, it is configurable by the user which can include in the generated code interesting features, and the reachability of each branch is assured (as we do for the validity of the IPMs).
\ins{The application of benchmarks is not limited only to pure software systems, but sometimes is applied even in systems embedding hardware. For example, benchmarks generated by exploiting machine learning are used to test computer networks~\cite{Cerquitelli2023}.}

\section{Conclusions}\label{sec:conclusions}

Testing and comparing combinatorial test generators is of paramount importance for the improvement, both in terms of performance and correctness, of the tools developed by practitioners in combinatorial testing.
However, this process requires the availability of a high number of benchmarks representing real-world examples and grasping all the aspects of interest.

For reducing this gap in evaluating test generators, in this paper, we have presented \tool, a generator of benchmark IPMs.
It is fully configurable by users, that can decide the type of parameters and constraints to be included in each model, their number and complexity, as well as the properties of the IPMs themselves (e.g., the ratios and the existence of at least one valid test case).

Its applicability has \chg{been already}{already been} demonstrated by its use during all the past editions of the CT Competition, held yearly during the International Workshop on Combinatorial Testing.
Moreover, in this paper, we have further extended the tool and unit-tested it by using a combinatorial test suite directly derived from its requirements.
As shown by the external validation activity, in which we have compared the IPMs available in the literature with those generable with \tool, we believe that our tool can be profitably used for evaluating test generators with synthetically generated benchmarks having the same characteristics as real-world systems.

\ins{As a future work, we may investigate approaches allowing \tool to solve the limitation of throwing away the generated IPM every time it is not solvable or compliant with the ratios requested by the user.
Moreover, we may include the generation of not solvable IPMs. 
This would allow users to test their generators not only when the model can be solved, but also in negative cases, and to verify that the generators are actually able to identify that condition.}


\bibliographystyle{plain}
\bibliography{bibliography}

\begin{thebibliography}{10}

\bibitem{Ansotegui2023}
Carlos Ansotegui and Eduard Torres.
\newblock A benchmark generator for combinatorial testing.
\newblock techreport, arxiv.org, 2023.

\bibitem{Bombarda2021}
Andrea Bombarda, Edoardo Crippa, and Angelo Gargantini.
\newblock An environment for benchmarking combinatorial test suite generators.
\newblock In {\em 2021 {IEEE} International Conference on Software Testing,
  Verification and Validation Workshops ({ICSTW})}, pages 48--56. {IEEE}, apr
  2021.

\bibitem{Bombarda2022a}
Andrea Bombarda and Angelo Gargantini.
\newblock Parallel test generation for combinatorial models based on
  multivalued decision diagrams.
\newblock In {\em 2022 {IEEE} International Conference on Software Testing,
  Verification and Validation Workshops ({ICSTW})}, pages 74--81. {IEEE}, apr
  2022.

\bibitem{Bombarda2023a}
Andrea Bombarda and Angelo Gargantini.
\newblock Incremental generation of combinatorial test suites starting from
  existing seed tests.
\newblock In {\em 2023 {IEEE} International Conference on Software Testing,
  Verification and Validation Workshops ({ICSTW})}. {IEEE}, April 2023.

\bibitem{Bombarda2023}
Andrea Bombarda, Angelo Gargantini, and Andrea Calvagna.
\newblock Multi-thread combinatorial test generation with smt solvers.
\newblock In {\em Proceedings of the 38th ACM/SIGAPP Symposium on Applied
  Computing}, SAC '23, New York, NY, USA, 2023. Association for Computing
  Machinery.

\bibitem{CTCompetition2022Repo}
Andrea Bombarda, Michael Wagner, and Manuel Leithner.
\newblock {CT-Competition 2022 page}.
\newblock
  \url{https://github.com/fmselab/CIT_Benchmark_Generator/tree/main/Benchmarks_CITCompetition_2022}.

\bibitem{CTCompetition2023Repo}
Andrea Bombarda, Michael Wagner, and Manuel Leithner.
\newblock {CT-Competition 2023 GitHub page}.
\newblock
  \url{https://github.com/fmselab/CIT_Benchmark_Generator/tree/main/Benchmarks_CITCompetition_2023}.

\bibitem{Cerquitelli2023}
Tania Cerquitelli, Michela Meo, Marilia Curado, Lea Skorin-Kapov, and
  Eirini~Eleni Tsiropoulou.
\newblock Machine learning empowered computer networks.
\newblock {\em Computer Networks}, 230:109807, July 2023.

\bibitem{Derks2023}
Christoph Derks, Daniel Strüber, and Thorsten Berger.
\newblock A benchmark generator framework for evolving variant-rich software.
\newblock {\em Journal of Systems and Software}, 203:111736, sep 2023.

\bibitem{Ferrer2011}
Javier Ferrer, Francisco Chicano, and Enrique Alba.
\newblock Benchmark generator for software testers.
\newblock In {\em {IFIP} Advances in Information and Communication Technology},
  pages 378--388. Springer Berlin Heidelberg, 2011.

\bibitem{Gargantini2018}
A.~Gargantini and M.~Radavelli.
\newblock Migrating combinatorial interaction test modeling and generation to
  the web.
\newblock In {\em 2018 IEEE International Conference on Software Testing,
  Verification and Validation Workshops (ICSTW)}, pages 308--317, April 2018.

\bibitem{Gargantini2014}
Angelo Gargantini and Paolo Vavassori.
\newblock Efficient combinatorial test generation based on multivalued decision
  diagrams.
\newblock In {\em Hardware and Software: Verification and Testing}, pages
  220--235. Springer International Publishing, 2014.

\bibitem{Garvin2009}
B.~J. {Garvin}, M.~B. {Cohen}, and M.~B. {Dwyer}.
\newblock An improved meta-heuristic search for constrained interaction
  testing.
\newblock In {\em 2009 1st International Symposium on Search Based Software
  Engineering}, pages 13--22, 2009.

\bibitem{Garvin2010}
Brady~J. Garvin, Myra~B. Cohen, and Matthew~B. Dwyer.
\newblock Evaluating improvements to a meta-heuristic search for constrained
  interaction testing.
\newblock {\em Empirical Software Engineering}, 16(1):61--102, July 2010.

\bibitem{Hasselbring2021}
Wilhelm Hasselbring.
\newblock Benchmarking as empirical standard in software engineering research.
\newblock In {\em Evaluation and Assessment in Software Engineering}. {ACM},
  June 2021.

\bibitem{Jin2020}
Hao Jin, Takashi Kitamura, Eun-Hye Choi, and Tatsuhiro Tsuchiya.
\newblock A comparative study on combinatorial and random testing for highly
  configurable systems.
\newblock In {\em Testing Software and Systems}, pages 302--309. Springer
  International Publishing, 2020.

\bibitem{Johansen2011}
Martin~Fagereng Johansen, {\O}ystein Haugen, and Franck Fleurey.
\newblock Properties of realistic feature models make combinatorial testing of
  product lines feasible.
\newblock In {\em Model Driven Engineering Languages and Systems}, pages
  638--652. Springer Berlin Heidelberg, 2011.

\bibitem{Karp1989a}
Richard~M Karp, Michael Luby, and Neal Madras.
\newblock Monte-carlo approximation algorithms for enumeration problems.
\newblock {\em Journal of Algorithms}, 10(3):429--448, sep 1989.

\bibitem{Khalsa2014}
Sunint~Kaur Khalsa and Yvan Labiche.
\newblock An orchestrated survey of available algorithms and tools for
  combinatorial testing.
\newblock In {\em 2014 {IEEE} 25th International Symposium on Software
  Reliability Engineering}, pages 324--334. {IEEE}, nov 2014.

\bibitem{Kuhn2004}
D.R. Kuhn, D.R. Wallace, and A.M. Gallo.
\newblock Software fault interactions and implications for software testing.
\newblock {\em IEEE Transactions on Software Engineering}, 30(6):418--421,
  2004.

\bibitem{PICTRepo}
{Microsoft Inc.}
\newblock {PICT GitHub page}.
\newblock \url{https://github.com/microsoft/pict}.

\bibitem{Nie2011}
Changhai Nie and Hareton Leung.
\newblock A survey of combinatorial testing.
\newblock {\em {ACM} Computing Surveys}, 43(2):1--29, jan 2011.

\bibitem{Niu2013}
Xintao Niu, Changhai Nie, Yu~Lei, and Alvin~T.S. Chan.
\newblock Identifying failure-inducing combinations using tuple relationship.
\newblock In {\em 2013 {IEEE} Sixth International Conference on Software
  Testing, Verification and Validation Workshops}. {IEEE}, March 2013.

\bibitem{petke2015practical}
Justyna Petke, Myra~B Cohen, Mark Harman, and Shin Yoo.
\newblock Practical combinatorial interaction testing: Empirical findings on
  efficiency and early fault detection.
\newblock {\em IEEE Transactions on Software Engineering}, 41(9):901--924,
  2015.

\bibitem{Segall2011}
Itai Segall, Rachel Tzoref-Brill, and Eitan Farchi.
\newblock Using binary decision diagrams for combinatorial test design.
\newblock In {\em Proceedings of the 2011 International Symposium on Software
  Testing and Analysis}. {ACM}, July 2011.

\bibitem{Thm2020}
Thomas Th\"{u}m.
\newblock A {BDD} for linux?
\newblock In {\em Proceedings of the 24th {ACM} Conference on Systems and
  Software Product Line: Volume A - Volume A}. {ACM}, October 2020.

\bibitem{Toda2016}
Takahisa Toda and Takehide Soh.
\newblock Implementing efficient all solutions {SAT} solvers.
\newblock {\em {ACM} Journal of Experimental Algorithmics}, 21:1--44, November
  2016.

\bibitem{TzorefBrill2018}
Rachel Tzoref-Brill and Shahar Maoz.
\newblock Modify, enhance, select: co-evolution of combinatorial models and
  test plans.
\newblock In {\em Proceedings of the 2018 26th {ACM} Joint Meeting on European
  Software Engineering Conference and Symposium on the Foundations of Software
  Engineering}. {ACM}, October 2018.

\bibitem{Wagner2020}
Michael Wagner, K.~Kleine, Dimitris Simos, R.~Kuhn, and R.~Kacker.
\newblock Cagen: A fast combinatorial test generation tool with support for
  constraints and higher-index.
\newblock In {\em International Workshop on Combinatorial Testing (IWCT 2020)},
  3 2020.

\bibitem{Younes2005}
Abdunnaser Younes, Paul Calamai, and Otman Basir.
\newblock Generalized benchmark generation for dynamic combinatorial problems.
\newblock In {\em Proceedings of the 7th annual workshop on Genetic and
  evolutionary computation}. {ACM}, June 2005.

\bibitem{Yu2013}
Linbin Yu, Yu~Lei, Raghu~N. Kacker, and D.~Richard Kuhn.
\newblock {ACTS}: A combinatorial test generation tool.
\newblock In {\em 2013 {IEEE} Sixth International Conference on Software
  Testing, Verification and Validation}, pages 370--375. {IEEE}, mar 2013.

\end{thebibliography}

\end{document}